\documentclass[11pt]{article}

\usepackage{amssymb,latexsym,amsmath}
\usepackage{xcolor}
\usepackage{bm}

\usepackage{hyperref}

\begin{document}

\newtheorem{theorem}{Theorem}[section]

\newtheorem{proposition}[theorem]{Proposition}

\newtheorem{lemma}[theorem]{Lemma}

\newtheorem{corollary}[theorem]{Corollary}

\newtheorem{definition}[theorem]{Definition}

\newtheorem{remark}[theorem]{Remark}

\newtheorem{exempl}{Example}[section]

\newenvironment{exemplu}{\begin{exempl}  \em}{\hfill $\surd$

\end{exempl}}

\newcommand{\ea}{\mbox{{\bf a}}}
\newcommand{\eu}{\mbox{{\bf u}}}
\newcommand{\ep}{\mbox{{\bf p}}}
\newcommand{\ed}{\mbox{{\bf d}}}
\newcommand{\eD}{\mbox{{\bf D}}}
\newcommand{\eK}{\mathbb{K}}
\newcommand{\eL}{\mathbb{L}}
\newcommand{\eB}{\mathbb{B}}
\newcommand{\ueu}{\underline{\eu}}
\newcommand{\ueo}{\overline{u}}
\newcommand{\oeu}{\overline{\eu}}
\newcommand{\ew}{\mbox{{\bf w}}}
\newcommand{\ef}{\mbox{{\bf f}}}
\newcommand{\eF}{\mbox{{\bf F}}}
\newcommand{\eC}{\mbox{{\bf C}}}
\newcommand{\en}{\mbox{{\bf n}}}
\newcommand{\eT}{\mbox{{\bf T}}}
\newcommand{\eV}{\mbox{{\bf V}}}
\newcommand{\eU}{\mbox{{\bf U}}}
\newcommand{\ev}{\mbox{{\bf v}}}
\newcommand{\eve}{\mbox{{\bf e}}}
\newcommand{\uev}{\underline{\ev}}
\newcommand{\eY}{\mbox{{\bf Y}}}
\newcommand{\eP}{\mbox{{\bf P}}}
\newcommand{\eS}{\mbox{{\bf S}}}
\newcommand{\eJ}{\mbox{{\bf J}}}
\newcommand{\leb}{{\cal L}^{n}}
\newcommand{\eI}{{\cal I}}
\newcommand{\eE}{{\cal E}}
\newcommand{\hen}{{\cal H}^{n-1}}
\newcommand{\eBV}{\mbox{{\bf BV}}}
\newcommand{\eA}{\mbox{{\bf A}}}
\newcommand{\eSBV}{\mbox{{\bf SBV}}}
\newcommand{\eBD}{\mbox{{\bf BD}}}
\newcommand{\eSBD}{\mbox{{\bf SBD}}}
\newcommand{\ecs}{\mbox{{\bf X}}}
\newcommand{\eg}{\mbox{{\bf g}}}
\newcommand{\paromega}{\partial \Omega}
\newcommand{\gau}{\Gamma_{u}}
\newcommand{\gaf}{\Gamma_{f}}
\newcommand{\sig}{{\bf \sigma}}
\newcommand{\gac}{\Gamma_{\mbox{{\bf c}}}}
\newcommand{\deu}{\dot{\eu}}
\newcommand{\dueu}{\underline{\deu}}
\newcommand{\dev}{\dot{\ev}}
\newcommand{\duev}{\underline{\dev}}
\newcommand{\weak}{\rightharpoonup}
\newcommand{\weakdown}{\rightharpoondown}
\renewcommand{\contentsname}{ }

\title{A symplectic Brezis-Ekeland-Nayroles principle}
\author{Marius Buliga\footnote{"Simion Stoilow" Institute of Mathematics of the Romanian Academy,
 PO BOX 1-764,014700 Bucharest, Romania, e-mail: Marius.Buliga@imar.ro } and 
G\'ery de Saxc\'e\footnote{ Laboratoire de M\'ecanique de Lille, UMR CNRS 8107, Universit\'e de Lille 1, Cit\'e scientifique, F59655 Villeneuve d'Ascq, France, e-mail: gery.desaxce@univ-lille1.fr}}
\date{This version: 12.08.2014}


\maketitle

\begin{abstract}
We propose a modification of the hamiltonian formalism which can be used for dissipative systems. This work continues \cite{bham} and advances by the introduction of a symplectic version of the Brezis-Ekeland-Nayroles principle \cite{Brezis Ekeland 1976} \cite{Nayroles 1976}. As an application we show how standard plasticity can be treated in our formalism.
\end{abstract}




\section{Introduction}

Realistic dynamical systems considered by engineers and physicists are subjected to energy loss. It may ensue from external actions, in which case we call them non conservative. On the other hand, if the cause is internal, resulting from a broad spectrum of phenomena such as collisions, surface friction, viscosity, plasticity, fracture, damage and so on, we name them dissipative. 

\vspace{.5cm}

However, at least in first approximation, the system can be considered as idealized and dissipation may often be neglected,  which allows for applying the methods of classical dynamics. The positions of the bodies are described by degrees of freedom $x^i$ which can be viewed as coordinates of a point $x$ in the system's configuration space $X$ and the motion is governed by a second order system of ODEs. A fruitful idea is to break the equations by considering the corresponding momenta $y_i$ as coordinates of a point $y$ living in the space $Y$, that leads to work in the phase space $X \times Y$ of points $z = (x, y)$. Hence the motion is governed by the first order system of canonical equations generated by the hamiltonian function $H$, allowing  the use of a very wide range of mathematical methods based on smooth functions to solve such problems. 

In modern presentation of this topic, differential geometry offers a powerful framework to tackle the problems from a global viewpoint. The main actors  are the symplectic form $\omega$ (an antisymmetric tensor of rank two or $2$-form) and the hamiltonian vector field $X H$ (the symplectic gradient of the hamiltonian).  Many such technics of classical dynamics are prototypes for infinite dimensional systems, for example ideal fluids of which the configuration space is the one of volume preserving diffeomorphisms. Among the other tools to solve dynamics problems, it is worth to quote the integral of motions and, in the modern language of Lie group theory, the momentum map linked to Noether's theorem.

\vspace{.5cm}

On the other hand, deformations of solids and motions of fluids are modeled through constitutive laws. Due to collisions, brittle fracture and threshold effects, most dissipative laws are non smooth and multivalued but experimental testing suggests that convexity is a keystone property of these phenomenological laws. In the framework of convex analysis, the usual gradient can be generalized thanks to the subdifferential $\partial \phi (z)$ (a set of subgradient of a convex and lower semicontinuous function $\phi$ at a given point $z$). Standard dissipative laws can be represented by a potential function $\phi$, convex but not differentiable everywhere, naturally leading to variational methods and unconstrained or constrained optimization problems with the numerical simulations in the background. Among them, Brezis-Ekeland-Nayroles principle \cite{Brezis Ekeland 1976} \cite{Nayroles 1976} is based on the time integration of the sum of dissipation potential $\phi$ and its Fenchel polar (analogous of Legendre polar for convex functions). Although not used much in the literature, the principle becomes noteworthy later because  it allows covering the whole evolution of the dissipative system at once.

\vspace{.5cm}

Classical dynamics is generally addressed through the world of smooth functions while the mechanics of dissipative systems deals with the one of non smooth functions. Unfortunately, both worlds widely ignore each other. The aim of this work is to lay strong foundations of bridging both worlds and their corresponding topics. In a previous paper \cite{bham}, the former author proposed the formalism of hamiltonian inclusions, able to model dynamical systems with $1$-homogeneous dissipation potential, (for laws such as brittle damage using Ambrosio-Tortorelli functional \cite{ambtor}). This formalism is a dynamical version of the quasistatic theory of rate-independent systems of Mielke.

The key-idea is to decompose additively the time rate $\dot{z}$ into reversible part $\dot{z}_R$ (the symplectic gradient) and dissipative or irreversible one $\dot{z}_I$, next to define the symplectic subdifferential $\partial^{\omega} \phi (z)$ of the dissipation potential. To get rid of the restrictive hypothesis of $1$-homogeneity (in particular to address viscoplasticity), we introduce in this work the symplectic Fenchel polar $\phi^{*\omega}$, then we  generalize the hamiltonian inclusion formalism by combining it with the  Brezis-Ekeland-Nayroles principle.

\textbf{Our aim is to build theoretical methods to model and analyze dynamical dissipative systems in a consistent geometrical frameworkm with the applications to  numerical approaches  not very far in the background}.  The objective is triple:
\begin{itemize}
\item \textit{Extending to the dissipative systems the geometrical methods of classical dynamics}.  The reversible part of the behaviour of the system (with respect to the additive decomposition of the time rate) is governed by a hamiltonian, therefore it can be studied by symplectic geometry techniques. The true dissipative part of the behaviour can be studied by convex analysis techniques. In this respect, the use of Lie group theory and the momentum map is valuable to analyze the symmetries of the problems and guess appropriate coordinates in which the problem is simpler from the point of view of the reversible part.
\item \textit{Exploring the dissipative rheological models in dynamical situations}. Such models are widely developed in statics but their identification in dynamics makes quick progress later  thanks to the improvements in experimental testing. Ee believe that the help of a consistent theoretical framework may be welcome for this task.
\item \textit{Using dynamical Brezis-Ekeland-Fenchel principle to solve evolution problems}. Indeed, step-by-step numerical methods  prevails today, but the weak point of this methods is that the errors are growing with the step number and the integration fails if it does not converge at a given step (difficulty to restart). On the contrary, the Brezis-Ekeland-Nayroles principle allows to have a consistent view of the whole evolution by determining simultaneously all the steps. Of course, solving space-time principles is more time-consuming but it could be improved by using model reduction methods such as  the Proper Generalized Decomposition (PGD) \cite{Ammar 2006}, \cite{Ammar 2007}, \cite{Bognet 2012}, \cite{Giner Chinesta 2013}.
\end{itemize}

Closer to the subject of this article, we cite the contributions of Aubin \cite{aubin2}, Aubin, Cellina and Nohel \cite{aubin}, Rockafellar \cite{rocka}, which considered various extensions of hamiltonian and lagrangian mechanics. In the article Bloch, Krishnaprasad, Marsden and Ratiu \cite{bloch} are explored hamiltonian systems with an added Rayleigh dissipation. A theory of quasistatic rate-independent systems is proposed by Mielke and Theil \cite{mielketh99}, Mielke \cite{mielke}, and developed towards applications in many papers, among them Mielke and Roub\'{\i}\v{c}ek \cite{MR06b}, see also the very recent Visintin \cite{visintin}.

Moreover, another advantage of Brezis-Ekeland-Nayroles principle is the easiness to be generalized. Indeed, it is worth to know that many realistic dissipative laws, called non-associated, cannot be cast in the mold of the standard ones deriving of a dissipation potential. To skirt this pitfall, the later author proposed in \cite{saxfeng} a new theory based on a function called bipotential. It represents physically the dissipation and generalizes the sum of the dissipation potential and its Fenchel polar, reason for which the  extension of the Brezis-Ekeland-Nayroles principle to bipotentials is natural and will be done in the future.

This eventual extension could be very beneficial, because bipotentials  applications to solid Mechanics are various: Coulomb's friction law \cite{sax CRAS 92}, non-associated Dr\"ucker-Prager  \cite{sax boussh IJMS 98} and Cam-Clay models \cite{Zouain 2010} in Soil Mechanics, cyclic Plasticity (\cite{sax CRAS 92},\cite{bodo sax EJM 01}) and Viscoplasticity \cite{hjiaj bodo CRAS 00} of metals with non linear kinematical hardening rule, Lemaitre's damage law \cite{bodo},  the coaxial laws (\cite{dangsax},\cite{vall leri CONST 05}). Such kind of materials are called implicit standard materials. A synthetic review of these laws can be found in the two later references. It is also worth to notice that monotone laws  which don't  admit a convex potential can be represented by Fitzpatrick's function \cite{Fitzpatrick 1988} which is a bipotential.

\paragraph{Acknowledgements.} This work has been done during Marius Buliga's visit at Laboratoire de M\'ecanique de Lille (CNRS mixed research laboratory 8107) supported by Universit\'e de Lille 1.

\section{Preliminaries and notations}

In this section we redefine the familiar notions of subdifferential, gradient and Fenchel transform, by using a symplectic form instead of the usual duality. 

We start with the following setting (but see Remark \ref{rem1} for possible generalizations). 
$X$ and $Y$ are topological, locally convex, real vector spaces of dual 
variables $x \in X$ and $y \in Y$. There is a duality product 
$$\langle \cdot , \cdot \rangle : X \times Y \rightarrow \mathbb{R}$$
 such that  any  continuous linear functional on $X$ (resp. on $Y$) has the form $x \mapsto \langle x,y\rangle$, for some $y \in Y$ (resp.  $y \mapsto \langle x,y\rangle$, for some  $x \in X$).

For a general element of $X \times Y$ we shall use the notation $z = (x,y)$, or
similar.  

The space $X \times Y$ has a natural symplectic  (i.e. bilinear and antisymmetric) form 
$\displaystyle \omega: \left( X \times Y \right)^{2} \rightarrow \mathbb{R}$ which is defined via the duality product by the formula:  for any $z=(x,y)$ and  $z'=(x',y')$   
$$\omega ( z, z' ) \, = \langle x , y' \rangle - \langle x' , y \rangle \quad
\quad . $$

\begin{definition} The symplectic subdifferential of $F: X \times Y \rightarrow \mathbb{R}\cup \left\{+\infty\right\}$, a convex lsc function, is the function which  associates to any 
$z = (x,y) \in X \times Y$ such that $F(z) < + \infty$ the set 
$$\displaystyle \partial^{\omega} F (z) \, = \, \left\{ 
z' \in X \times Y \mbox{ : } \forall \, z" \in X \times Y \quad  
F(z+z") \,  \geq \,  F(z) \, + \, 
\omega (z' , z") \right\} $$ 
\label{dssub}
\end{definition}

In \cite{bham} Definition 2.2, the symplectic subdifferential is denoted by $XF$, here we use a different notation. 

\begin{definition}
The symplectic polar, or the symplectic Fenchel transform of $F: X \times Y \rightarrow \mathbb{R}\cup \left\{+\infty\right\}$, a convex lsc function, is the function: 
$$F^{*\omega}(z') \, = \, \sup \left\{ \omega(z',z) - F(z) \mbox{ : } z \in X \times Y \right\}$$
\label{dspolar}
\end{definition}

\paragraph{The particular case $X=Y$.} In this case the duality  $\displaystyle \langle \cdot, \cdot \rangle$ is a scalar product.  Moreover the space $X \times Y$ is dual with itself, with the duality product: 
$$\langle \langle (x,y), (x', y') \rangle \rangle \, = \, \langle x, x' \rangle + \langle y, y' \rangle \quad . $$ 
We  introduce now  the linear function 
$$J: X \times Y \rightarrow X \times Y \, \, , \, \, J(x,y) = (-y,x)$$ 
and we remark that we  have  the relations $\displaystyle J^{2} = - I$ and 
$$\omega((x,y), (x',y')) \, = \, \langle \langle J(x,y), (x',y') \rangle \rangle $$
$$\omega( -J z', z") \, = \, \langle \langle z', z" \rangle \rangle \quad .$$
Notice that  $J$ makes no sense in the general case when $X \not = Y$.

The subdifferential of $\displaystyle F: X \times Y \rightarrow \mathbb{R} \cup \left\{ + \infty \right\}$ is by definition: 
$$\partial F(z) \, = \, \left\{ z' \in X \times Y \mbox{ : }  \forall \, z" \in X \times Y \quad  
F(z+z") \,  \geq \,  F(z) \, + \, \right. $$ 
$$\left. + \,  \langle \langle z' , z" \rangle \rangle \right\} \quad . $$
By comparison with the definition \ref{dssub} of the symplectic subdifferential of $F$, we obtain: 
$$z' \in \partial^{\omega} F(z) \, \Leftrightarrow \, Jz' \in \partial F(z) \quad ,$$
$$z' \in \partial F(z) \, \Leftrightarrow \, -Jz' \in \partial^{\omega} F(z) \quad ,$$
in the particular case $X = Y$.

We continue by writing the  definition of the Fenchel conjugate of a function $\displaystyle F: X \times Y \rightarrow \mathbb{R} \cup \left\{ + \infty \right\}$: 
$$F^{*}(z) \, = \, \sup \left\{ \langle \langle z', z \rangle \rangle \, - \, F(z') \mbox{ : } z' \in X \times Y \right\} \quad . $$ 
Remark that, from definition \ref{dspolar} of the symplectic polar of $F$, we get: 
$$F^{*\omega}(z) \, = \, F^{*}(Jz) \quad , $$
in the particular case $X =Y$.

Let's go back to the general setting, when we don't suppose that $X = Y$. (Then, of course, the function $J$ no longer makes sense.) In this generality  we have the following symplectic version of the Fenchel inequality. 

\begin{theorem}
Suppose that $\phi$ is convex, lsc. Then for any $z,z' \in X \times Y$ we have 
$$\phi(z)\, + \, \phi^{*\omega} (z')  \, \geq \, \omega(z',z)$$ 
and the equality is achieved if and only if $\displaystyle z' \in \partial^{\omega} \phi (z)$. 
\label{sfenchelthm}
\end{theorem}

\begin{remark}
In the case $X=Y$ this is just a reformulation of the usual Fenchel inequality. Indeed, in this case we may replace $\displaystyle \phi^{*\omega} (z')$ by $\displaystyle \phi^{*} (Jz')$ and $\displaystyle \omega(z',z)$ by $\displaystyle \langle \langle Jz', z \rangle \rangle$. 
\label{rem33}
\end{remark}

\paragraph{Proof.}  Taking inspiration from  the remark \ref{rem33}, we notice that we can reproduce verbatim the proof of the usual Fenchel inequality, even if we place ourselves in the general case. 

By the Definition \ref{dspolar} of the symplectic polar we have that for any $\displaystyle z, z' \in X \times Y$  
$$ \phi^{*\omega}(z') \, \geq \,  \omega(z',z) - \phi(z)$$ 
which gives the symplectic Fenchel inequality. The equality is attained if and only if $\displaystyle \phi(z)$ and $\displaystyle \phi^{*\omega}(z')$ are finite and moreover: 
$$\phi^{*\omega}(z') \, = \, \omega(z',z) - \phi(z) \, \geq \, \omega(z', z + z") - \phi(z+z")$$ 
for any $\displaystyle z" \in X \times Y$. But this is equivalent with: $\displaystyle \phi(z)$ and $\displaystyle \phi^{* \omega}(z')$ are finite and moreover 
$$\phi(z+z") \, \geq \, \phi(z) + \omega(z', z")$$ 
for any $\displaystyle z" \in X \times Y$, i.e. with $\displaystyle z' \in \partial^{\omega} \phi (z)$. \hfill $\square$

\begin{remark} 
We may replace the pair $X, Y$ and the associated symplectic form $\omega$ by a symplectic manifold $(M, \omega)$. In this case $\omega$ becomes a field of antisymmetric bilinear forms, i.e. 
$$\displaystyle \omega = \omega_{z}: T_{z} M \times T_{z} M \rightarrow \mathbb{R}$$ 
for any $z \in M$. Here we denote by $\displaystyle T_{z}M$ the tangent space to $M$ at $z$. 
We consider then functions 
$$\displaystyle F: TM \rightarrow \mathbb{R} \cup \left\{ + \infty \right\}$$
where $TM$ denotes the tangent bundle of $M$. We ask that the function $F$ is lsc, and moreover for any $z \in M$ the function $\displaystyle Z \in T_{z}M \mapsto F(z,Z)$ is convex. 

In this more general setting, the definition of the subdifferential of a function $F$ is 
the set $\displaystyle \partial^{\omega} F (z,Z) \subset T_{z}M$ given  by 
$$\displaystyle \partial^{\omega} F (z, Z) \, = \, \left\{ 
Z' \in T_{z}M \mbox{ : } \forall \, Z" \in T_{z}M \quad  
F(z,Z+Z") \,  \geq \,  F(z,Z) \, + \, 
\omega_{z} (Z' , Z" \right\} $$ 
and the definition of the symplectic polar changes to: for any $z \in M$ and any $\displaystyle Z' \in T_{z}M $ 
$$F^{*\omega}(z,Z') \, = \, \sup \left\{ \omega_{z}(Z',Z) - F(z,Z) \mbox{ : } z \in T_{z}M \right\} \quad .$$
Finally, the symplectic Fenchel inequality takes the form: for any $\displaystyle z \in M$ and for any $\displaystyle Z, Z' \in T_{z}M$ we have 
$$\phi(z,Z)\, + \, \phi^{*\omega} (z,Z')  \, \geq \, \omega_{z}(Z',Z)$$ 
and the equality is realized if and only if $\displaystyle Z' \in \partial^{\omega} \phi (z,Z)$. 
\label{rem1}
\end{remark}

\section{Hamiltonian evolution}

\begin{definition}
A function $F: X \times Y \rightarrow \mathbb{R} \cup \left\{+\infty \right\}$ has a symplectic gradient in a point $z = (x,y)$ if $F(x,y) < + \infty$ and there exists $\displaystyle XF(x,y) = (u,v) \in X \times Y$, called the symplectic gradient of $F$ in $(x,y)$,  such that 
\begin{enumerate}
\item[(a)]  for all $y' \in Y$ we have 
$$\lim_{\varepsilon \rightarrow 0} \frac{1}{\varepsilon} \left[ F(x, y +
\varepsilon y') - F(x,y) \right] \, = \, \langle u , y' \rangle $$
\item[(b)]  for all $x' \in X$ we have 
$$\lim_{\varepsilon \rightarrow 0} \frac{1}{\varepsilon} \left[ F(x +
\varepsilon x', y ) - F(x,y) \right] \, = \, -  \langle x', v \rangle $$
\end{enumerate}
\label{dsg}
\end{definition}

Let us denote by $DF(x,y) \in X \times Y$ the vector defined by: for any $(x',y') \in X \times Y$ 

$$\lim_{\varepsilon \rightarrow 0} \frac{1}{\varepsilon} \left[ F(x+ \varepsilon x', y +
\varepsilon y') - F(x,y) \right] \, = \, \langle \langle DF(x,y), (x',y') \rangle \rangle  \quad .$$

Then, in the  particular case  $X = Y$, by denoting $z = (x,y)$, we have  $\displaystyle J X F (z) = DF(z)$ and $\displaystyle X F (z) = - J DF(z)$.  Moreover, we have the following useful formula, obtained from the ones about the symplectic subdifferential in the case $X = Y$: 
$$XF(z) \in \partial^{\omega} F(z) \, \Leftrightarrow \, DF(z) \in \partial F(z) \quad ,$$
Moreover, suppose that we use a function $\displaystyle F = F(t,z)$ defined over $\displaystyle \mathbb{R} \times X \times Y$, where we see $\displaystyle X \times Y$ as the symplectic manifold $M$, like in the Remark \ref{rem1}. Then, from the formula 
$$\omega(XF(t,z), Z') \, = \, \langle \langle D_{z}F(t,z), Z' \rangle \rangle$$ 
we obtain the following: for any smooth curve $t \mapsto z(t) \in X \times Y$ we denote by 
$\displaystyle \dot{z}(t)$ the derivative of the curve with respect to $t$ and we have 
\begin{equation}
\frac{d}{dt} F(t,z(t)) \, = \, D_{t} F(t,z(t)) \, + \,  \omega(XF(t,z), \dot{z}(t))  \quad . 
\label{neededdt}
\end{equation}

\begin{definition}
Given a function $H = H(t,x,y) = H(t,z)$ called the hamiltonian, a curve $z: [0,T] \rightarrow X \times Y$ is an evolution curve of that hamiltonian function if it satisfies the  equation: 
\begin{equation}
\dot{z}(t) \,  = \, X H(t, z(t)) \quad .
\label{hevo}
\end{equation} 
\label{dhamev}
\end{definition}

Let's posit in the case $X = Y$ and consider the following example of a hamiltonian (further we use the notation $z = (q,p) \in X \times Y$, with the interpretation that $q$ is a position and $p$ is a momentum): 
$$H(t,z) \, = \, H(t,q,p) \, = \, \frac{1}{2m} \langle p, p \rangle^{2} \, + \, W(q) \, - \, \langle f(t) , q \rangle \quad .$$
Here $m$ has the interpretation of mass, $W$ is an energy and $f(t)$ is an external force. 
The equation (\ref{hevo}) is then 
$$(\dot{q}, \dot{p}) \, = \, (D_{p} H (t,q,p), -D_{q}H(t,q,p))$$ 
which is equivalent with the following system
$$\left\{ \begin{array}{l}
\dot{q} \, = \, \frac{1}{m} p \\
\dot{p} \, = \, - D_{q} W(q) +  f(t)
\end{array}
\right. $$
which is the same as 
$$m \ddot{q} \, = \, - D_{q} W(q) + f(t) \quad . $$
This is the Euler-Lagrange equation for the lagrangian function 
$$L(t,q, \dot{q}) \, = \, \frac{1}{2} m \langle \dot{q} , \dot{q} \rangle \, - \, W(q) \, +  \,  \langle f(t) , q \rangle $$
which describes a mechanical system with position $q$, kinetic energy $\displaystyle \frac{1}{2} m \langle \dot{q} , \dot{q} \rangle$, potential energy $W(q)$, subjected to external forces $f(t)$. 

If we stay on the Hamiltonian side, we remark that the system does not dissipate, in the sense that for $z(t) = (q(t), p(t))$ a solution of the equation (\ref{hevo}), we have 
$$ 0 \, = \, \omega(XH(t,z(t)), \dot{z}(t)) \, = \, \frac{d}{dt} H(t,z(t)) \, - \, D_{t} H(t,z(t)) \quad . $$
For example, if $\displaystyle D_{t} H(t,z(t)) = 0$ (as it is the case if $H = H(z)$ only) then the hamiltonian $H$ is constant along any evolution curve.

If we look at the lagrangian side then we remark that a complete model needs also equations which describe the dissipative behaviour of the system.

We shall need later the following definition of the Poisson bracket. 

\begin{definition}

Let $\displaystyle Der(X,Y)$ be the linear space of  functions 
$f: X\times Y \rightarrow \mathbb{R}$ which have a continuous symplectic gradient. 

The Poisson bracket is the bilinear, antisymmetric form 
$$\displaystyle \left\{ \cdot , \cdot
\right\} : Der(X,Y) \times Der(X,Y) \rightarrow \mathbb{R}^{X\times Y}$$ 
defined by: 
$\displaystyle \left\{ f , g \right\} \, = \, \omega \left( X_f , X_g \right)$. 
\label{defobj}
\end{definition}

\section{The symplectic BEN principle}

In this section we propose a modification of the hamiltonian formalism with the effect that it can apply to dissipative systems. This proposal builds further from the initial one of hamiltonian systems with convex dissipation introduced in \cite{bham}.

A hamiltonian $H = H(t,x,y) = H(t,z)$ and a dissipation potential $\displaystyle \phi= \phi( \dot{z})$,  
$$\phi:  X \times Y \rightarrow \mathbb{R} \cup \left\{ + \infty \right\}$$ are given. We shall suppose that the dissipation potential $\phi$ is convex and lsc. 

For any evolution curve $z: [0,T] \rightarrow X \times Y$,  we consider the additive decomposition of $\displaystyle \dot{z}$  into "reversible" and "irreversible" parts:
\begin{equation}
\dot{z} = \dot{z}_{R} + \dot{z}_{I} \quad , \quad   \dot{z}_{R} = X H (z) \, \, , \, \, \dot{z}_{I} = \dot{z} - XH(z) \quad . 
\label{decompo}
\end{equation}
Alternative good names would be "conservative" instead of "reversible" and "non conservative" instead of "irreversible". 

If the evolution is hamiltonian, with $H$ as the Hamiltonian function, i.e. $\displaystyle \dot{z} = X H (z)$, then $\displaystyle \dot{z}_{I} = 0$ and $\displaystyle \dot{z}_{R} = \dot{z}$. 

We propose the following principle (definition). Further "BEN" is a short notation for "Brezis-Ekeland-Nayroles". 

\begin{definition} (The symplectic BEN principle.) An evolution curve  $t \in [0,T] \mapsto z(t) \in X \times Y$ satisfies the symplectic Brezis-Ekeland-Nayroles principle for the hamiltonian $H$ and dissipation potential $\phi$ if for almost any $t \in [0,T]$ we have 
\begin{equation}
\phi(\dot{z}(t)) + \phi^{*\omega}( \dot{z}_{I}(t) ) = \omega(\dot{z}_{I}(t),\dot{z}(t)) \quad . 
\label{sben1}
\end{equation}
\label{dsben1}
\end{definition}

\begin{remark}
We use the name "Brezis-Ekeland-Nayroles" principle because, as explained in the section 
\ref{ssplas}, relation \ref{sbenp3}, in the particular case of standard plasticity the symplectic BEN principle reduces to the Brezis-Ekeland-Nayroles principle \cite{Brezis Ekeland 1976} \cite{Nayroles 1976} if we neglect the dynamical terms. 
\label{remwben}
\end{remark}

Two other, equivalent forms of the symplectic BEN principle are given in the next proposition. 

\begin{proposition}
 An evolution curve  $t \in [0,T] \mapsto z(t) \in X \times Y$ satisfies the symplectic Brezis-Ekeland-Nayroles principle for the hamiltonian $H$ and dissipation potential $\phi$ if and only if it satisfies one of the following: 
\begin{enumerate}
\item[(a)] for almost  every $t \in [0,T]$ 
\begin{equation}
\dot{z}(t) - X H(t,z(t) ) \, \in \, \partial^{\omega} \phi(\dot{z}) \quad . 
\label{sben3}
\end{equation}
\item[(b)] the evolution curve minimizes the functional 
\begin{equation}
\Pi(z') = \int_{0}^{T} \left\{ \phi(\dot{z'}(t)) + \phi^{*\omega}(\dot{z}'_{I}(t)) -  \frac{\partial H}{\partial t}(t,z'(t)) \right\} \mbox{ dt} \,  + 
\label{sben2}
\end{equation}
$$ + \,  H(T, z'(T)) $$
among all curves $z': [0,T] \rightarrow X \times Y$ such that $\displaystyle z'(0) = z(0)$. 
\end{enumerate}
\label{pequisben}
\end{proposition}

\paragraph{Proof.} (a)  We apply Theorem \ref{sfenchelthm} to (\ref{sben1}) and we obtain that  
$$\dot{z}_{I}(t) \in \partial^{\omega} \phi( \dot{z}(t)) \quad . $$
But this is the same as (\ref{sben3}), by using the decomposition (\ref{decompo}). 

(b) The formulation (\ref{sben2}) is obtained by integration on $[0,T]$ of the (\ref{sben1}). Indeed, remark that from (\ref{sben1}) we get: 
$$\int_{0}^{T} \left\{ \phi(\dot{z'}(t)) + \phi^{*\omega}(\dot{z'}_{I}(t)) - \omega(\dot{z'}_{I}(t),\dot{z'}_{R}(t)) \right\} \mbox{ dt} = 0 \quad . $$
But we know from the symplectic Fenchel inequality that the integrand is always non-negative, for any evolution curve, therefore any solution of the (\ref{sben1}) which satisfies the initial condition  $\displaystyle z'(0) = z(0)$ makes the integral equal to 0, thus being a minimizer of the integral. A short computation shows that for any curve $z'$ which satisfies the initial condition we have 
$$ \int_{0}^{T} \left\{ \phi(\dot{z'}(t)) + \phi^{*\omega}(\dot{z'}_{I}(t)) - \omega(\dot{z'}_{I}(t), \dot{z'}_{R}(t)) \right\} \mbox{ dt} = \Pi(z')  -  H(0,z(0))$$
therefore any solution of (\ref{sben1}) which satisfies the initial condition is also a solution of (\ref{sben2}). Conversely, any solution of (\ref{sben2}) satisfies the initial condition and it satisfies (\ref{sben1}) for almost every $t \in [0,T]$. \hfill $\square$

\begin{remark}
The symplectic BEN principle in the form (\ref{sben3}) appears in \cite{bham} Definition 2.3, relation (14). The notation is slightly different, in the mentioned reference is used a "dissipation function" $\mathcal{R}$, which corresponds to the dissipation potential $\phi$ from here. Notice however that in this article we don't take as a hypothesis that $\phi$ is a non negative function, like in \cite{bham} Definition 2.3 (a). With the formulation proposed here, there is a different possible positivity hypothesis which we might add.  Indeed, because of the antisymmetry of the symplectic form, when we use the relation (\ref{neededdt}) for the function $H$ we get: 
$$ \omega(\dot{z}_{I},\dot{z}) = - \omega (X H, \dot{z}) 
              = - \frac{d}{dt} \left[ H(t,z(t)) \right] + \frac{\partial H}{\partial t} ( t, z(t)) \quad .
$$
From the symplectic BEN principle (\ref{sben11}) we obtain  the following energy balance
$$ \phi(z, \dot{z}) + \left( \phi\left(z, \cdot\right) \right)^{*\omega}(\dot{z}_{I})\,  =  \,  - \frac{d}{dt} \left[ H(t,z(t)) \right] + \frac{\partial H}{\partial t} ( t, z(t)) \quad . $$
If the system dissipates then we expect that the quantity from the right hand side of the previous equation is non negative. A sufficient condition is  that $\phi$ satisfies:  any $z, z'\in X\times Y $
$$ \phi(z) + \phi^{*\omega}(z') \geq 0  \quad .
$$
In the particular case when $\phi$ is positively 1-homogeneous (as it happens for many interesting physical systems) then $\displaystyle  \phi^{*\omega}$ takes only the values $\displaystyle 0, +\infty$. Then, for positively 1-homogeneous dissipation $\phi$ we have 
$$\mbox{ if } \phi^{*\omega}(z') < + \infty \quad \mbox{ then } \phi(z) +  \phi^{*\omega}(z') \, = \, \phi(z) \quad .$$
Therefore in this case the positivity of $\phi(z)$ is equivalent with the positivity of 
$\displaystyle    \phi(z) + \phi^{*\omega}(z')$. 
\label{remcomp}
\end{remark}

We may generalize, in the setting explained in the Remark \ref{rem1},  the three variants of the symplectic BEN principle by making $\displaystyle \phi = \phi(z,\dot{z})$ and writing, for example instead of the (\ref{sben1}) the following: 

\begin{equation}
\phi(z, \dot{z}) + \left( \phi\left(z, \cdot\right) \right)^{*\omega}(\dot{z}_{I}) = \omega(\dot{z}_{I},\dot{z}_{R}) \quad . 
\label{sben11}
\end{equation}

\paragraph{Invariance of the symplectic BEN principle.}  The symplectic BEN principle (\ref{sben11}) or its equivalent formulation (\ref{sben3}) are stable under reparameterization. This means the following. Mathematically the dissipation potential $\phi$ is a function defined on the tangent bundle of $X \times Y$. Consider then a reparameterization $\displaystyle (x',y') = \Psi(x,y)$ which preserves the symplectic form, in the sense that the function $\Psi: X \times Y \rightarrow X \times Y$ is bijective and differentiable everywhere, and that if we denote by $\displaystyle D\Psi(x,y) \in Lin(X \times Y, X \times Y)$ it's derivative at $(x,y)$ then for any pair of "tangent vectors" $\displaystyle (x_{1}, y_{1}) , (x_{2}, y_{2}) \in X \times Y$ we have: 
$$ \omega( D \Psi(x,y)(x_{1}, y_{1}) , D\Psi(x,y) (x_{2}, y_{2})) \, = \, \omega( (x_{1}, y_{1}), (x_{2}, y_{2})) \quad . $$
Then, in the new coordinates $(x',y')$ the dissipation potential transforms as: 
$$\phi'(\Psi(x,y), D \Psi(x,y)(x_{1},y_{1})) \, = \, \phi((x,y), (x_{1},y_{1})) \quad .$$
The hamiltonian $H$ becomes $H'(t,\Psi(x,y)) = H(t,x,y))$. 

The symplectic polar of $\displaystyle \phi( (x,y), \cdot)$ will be, in the new coordinates: 
$$\left( \phi'\left(\Psi(x,y),  \cdot \right) \right)^{*\omega} (D \Psi(x,y)(x_{1}, y_{1})) \, = \, $$ 
$$= \, \sup  \left\{ \omega( D \Psi(x,y)(x_{1}, y_{1}), (x',y'))   \, - \, \phi(\Psi(x,y), (x',y')) \mbox{ : } (x',y')  \in X \times Y \right\} \, =   $$ 
$$ = \,  \sup  \left\{ \omega( D \Psi(x,y)(x_{1}, y_{1}), D \Psi(x,y)(x_{2}, y_{2}))   \, - \, \phi'(\Psi(x,y), D \Psi(x,y)(x_{2}, y_{2})) \right. $$ $$ \left.  \mbox{ : } (x_{2},y_{2})  \in X \times Y \right\} \, = $$
$$= \, \sup \left\{ \omega( (x_{1},y_{1}), (x_{2},y_{2})   \, - \, \phi((x,y),(x_{2},y_{2})) \mbox{ : }  (x_{2},y_{2})  \in X \times Y \right\}  \, =$$  $$ = \, \left( \phi\left((x,y), \cdot \right) \right)^{*\omega} ((x_{1}, y_{1})) \quad . $$

In the new coordinates the curve $\displaystyle z(t) = (x(t),y(t))$ becomes $z'(t) = \Psi(x(t),y(t))$, with the associated velocity 
$$\dot{z'} (t) \, = \, D \Psi(x(t),y(t)) \dot{z}(t) \quad . $$

Therefore the  symplectic BEN principle (\ref{sben11}) expressed in the new coordinates, with $\phi'$ and $H'$ instead of $\phi$ and $H$, will be the same as the original one.

Let us explore further the consequences of the symplectic BEN principle. 

\begin{proposition}
The curve $z: [0,T] \rightarrow X \times Y$ is a solution of the symplectic BEN principle (\ref{sben11}) if and only if for almost every $t \in [0,T]$ and  for any function $f: [0,T] \rightarrow  Der(X \times Y)$, $f = f(t, z)$  which is derivable with respect to $t$  we have: 
\begin{equation}
\phi(z(t), \dot{z}(t) - X f(t,z(t))) \, \geq \, \frac{d}{dt} \left[ f(t,z(t)) \right] - \frac{\partial f}{\partial t} ( t, z(t)) +  
\label{for1}
\end{equation}
$$+ \omega(XH(t,z(t)), Xf(t,z(t))) + \phi(z(t), \dot{z}(t)) \quad .$$
Moreover, if the curve $z$ satisfies (\ref{for1}) for any $f \in Der(X\times Y)$ (i.e. $f$ not depending on time) then $z$ is a solution of the symplectic BEN principle (\ref{sben11}). 
\label{pfor1}
\end{proposition}

\paragraph{Proof.}
We know that (\ref{sben11}) is equivalent with (\ref{sben3}), more precisely we know that $z$ satisfies (\ref{sben11}) if and only if it satisfies the following:
\begin{equation}
\dot{z} - X H(t,z(t)) \, \in \, \partial^{\omega} \phi(z(t), \cdot)  (\dot{z}) \quad . 
\label{sben33}
\end{equation}
The relation  (\ref{sben33}) means that for any $t \in [0,T]$ and for any $z' \in X \times Y$ we have: 
\begin{equation}
\phi(z(t), \dot{z}(t) + z') \, \geq \, \omega(\dot{z}(t) - XH(t,z(t)), z') + \phi(z(t), \dot{z}(t)) \quad . 
\label{inter1}
\end{equation}
Let us choose now $z' = - Xf(t,z(t))$. We get then 
\begin{equation}
\phi(z(t), \dot{z}(t) - X f(t,z(t))) \, \geq \, \omega(Xf(t,z(t)), \dot{z}(t)) + 
\label{inter2}
\end{equation}
$$+ \omega(XH(t,z(t)), Xf(t,z(t))) + \phi(z(t), \dot{z}(t))  \quad . $$
But this is the same as (\ref{for1}) because 
$$ \omega(Xf(t,z(t)), \dot{z}(t)) \, = \,  \frac{d}{dt} \left[ f(t,z(t)) \right] - \frac{\partial f}{\partial t} ( t, z(t)) $$ 
by the definition of the symplectic gradient. 

For the converse implication, for any $z' \in X \times Y$ let's pick the function $f(z) = \omega(z,z')$. Then $Xf(z) = - z'$ for any $z \in X \times Y$, therefore we can trace back our steps from (\ref{for1}) applied for this choice of $f$ to (\ref{inter1}). Because this can be done for any $z' \in X \times Y$ it follows that (\ref{inter1}) is true for any $z' \in X \times Y$, which proves that the curve $z$ satisfies the symplectic BEN principle (\ref{sben33}). \hfill $\square$

\begin{proposition}
Suppose that the curve $z: [0,T] \rightarrow X \times Y$ is a solution of the symplectic BEN principle (\ref{sben11}). Then for almost every $t \in [0,T]$ we have 
\begin{equation}
\phi(z(t), \dot{z}(t) - X f(t,z(t))) \, \geq \, \omega(X f(t, z(t)), \dot{z}(t)) + \phi(z(t), \dot{z}(t)) 
\label{b1}
\end{equation}
for any function $f$ with the property that 
$$\omega(Xf(t,z(t)) , X H (t,z(t))) = 0 \quad , $$ 
i.e. for any integral of motion of the hamiltonian $H$. 
\label{pb1}
\end{proposition}

\paragraph{Proof.} The hypothesis implies that $z$ and $f$ satisfy (\ref{inter2}). We use then the fact that $f$ is an integral of motion of the hamiltonian $H$ in order to get (\ref{b1}). \hfill $\square$

\begin{proposition}
Suppose that the curve $z: [0,T] \rightarrow X \times Y$  is a solution of the symplectic BEN principle (\ref{sben11}). Then for any function $f: [0,T] \rightarrow  Der(X \times Y)$, $f = f(t, x,y)$  which is derivable with respect to $t$  we have: 
\begin{equation}
\int_{0}^{T} \left[ \phi(z(t), \dot{z}(t) - X f(t,z(t)))  - \phi(z(t),\dot{z}(t) \right] \mbox{ d}t  \, 
\label{for2}
\end{equation}
$$\geq \, f(T,z(T)) - f(0,z(0)) \, + \, \int_{0}^{T} \left[   \left\{H,f\right\} (t,z(t)) -  \frac{\partial f}{\partial t} ( t, z(t)) \right] \mbox{ d}t  \quad . $$
Conversely, any solution of the problem (\ref{for2}) is also a solution of the symplectic BEN principle (\ref{sben11}) for almost every $t \in [0,T]$. 

\label{pfor2}
\end{proposition}

\paragraph{Proof.} 
By integration with respect to time of (\ref{for1}), then by integration by parts  we obtain (\ref{for2}). Conversely, we may pick the "test functions" $f$ to be with compact support with respect to the time variable, which gives the second claim. \hfill $\square$

\begin{proposition}
Any solution of the problem (\ref{for2}) satisfies the dissipation balance equation: for every $\tau \in [0,T]$ 
\begin{equation}
\int_{0}^{\tau} \left[ \phi(z(t), \dot{z}(t)) + \left(\phi(z(t), \cdot) \right)^{*\omega}(\dot{z}(t) - X H(t,z(t))) \right] \mbox{ d}t \, = 
\label{for3}
\end{equation}
$$= \, H(0, z(0)) - H(\tau, z(\tau)) \, + \, \int_{0}^{\tau} \frac{\partial H}{\partial t} (t, z(t)) \mbox{ d}t  $$
and the dissipation inequality: 
\begin{equation}
\phi(z(t), \dot{z}_{I}(t)) \, \geq \,  \frac{d}{dt} \left[ H(t,z(t)) \right] - \frac{\partial H}{\partial t} ( t, z(t)) + \phi(z(t), \dot{z}(t)) \quad . 
\label{b2}
\end{equation}
\label{pfor3}
\end{proposition}

\paragraph{Proof.} Indeed, solutions of (\ref{for2}) satisfy for almost every $t \in[0,T]$ the symplectic BEN principle (\ref{sben11}). We integrate it from $0$ to $\tau$ and we obtain the dissipation balance equation. For the inequality (\ref{b2})  we use Proposition \ref{pb1} for $f = H$.  \hfill $\square$  

It is interesting to notice that the dissipation balance equation (\ref{for3}) is a generalization of \cite{bham} Theorem 2.7, relation (21). The mentioned theorem has among the hypotheses that the dissipation potential is 1-homogeneous. Here we don't need this positivity hypothesis. The relation (\ref{for3}) is a generalization of \cite{bham} relation (21) because in the case when the dissipation $\phi$ is 1-homogeneous then, as explained in Remark \ref{remcomp}, we have $\displaystyle \phi(z) +  \phi^{*\omega}(z') \, = \, \phi(z)$ whenever the quantity from the left hand side is finite. 

\section{Application: standard plasticity}
\label{ssplas}

We would like now to illustrate the general formalism and to show how it allows to develop powerful variational principles for dissipative systems within the frame of continuum mechanics. To begin with, we tackle the standard plasticity in small deformations based on the additive decomposition of strains into reversible and irreversible strains:
$$ \bm{\varepsilon} = \bm{\varepsilon}_R + \bm{\varepsilon}_I
$$ 
where $\bm{\varepsilon}_I$ is a plastic strain $\bm{\varepsilon}_p$. 

The present modeling can immediately be extended to similar constitutive laws by considering alternatively viscous or viscoplastic strains depending on the material behaviour.

Let $\displaystyle \Omega \subset \mathbb{R}^{n}$ be a bounded, open set, with piecewise smooth boundary $\partial \Omega$.  The elements of the space $X$ are fields $x = (\bm{u},\bm{\varepsilon}_I) \in U \times E$ where $\bm{\varepsilon}_I$ is the irreversible strain field and $\bm{u}$ is a displacement field on the body $\Omega$ with trace $\bar{\bm{u}}$ on $\partial \Omega$.  The elements of the  corresponding dual space  $Y$  are  of the form  $y = (\bm{p},\bm{\pi})$. Unlike $\bm{p}$ which is clearly the linear momentum, we do not know at this stage the physical meaning of $\bm{\pi}$. We denote by $z = (x,y)$.

The duality between  the spaces $X$ and $Y$ has the form 
$$\langle x, y \rangle \, = \, \int_{\Omega} \left(  \langle \bm{u},\bm{p} \rangle + \langle \bm{\varepsilon}_I , \bm{\pi} 
\rangle \right) \mbox{ d}x \quad , $$ 
where the duality products which appear in the integral are finite dimensional duality products on the image of the fields $\bm{u}, \bm{p}$ (for our example this means a scalar product on $\displaystyle \mathbb{R}^{3}$) and on the image of the fields $\bm{\varepsilon}, \bm{\pi}$ (in this case this is a scalar product on the space of 3 by 3 symmetric matrices). We denote all these standard dualities by the same $\langle \cdot , \cdot \rangle$ symbols. 

The total hamiltonian of the structure is taken of the integral  form
$$ H (t, z) = \int_{\Omega} \left\lbrace \dfrac{1}{2 \rho} \parallel \bm{p} \parallel^ 2 
                           + w(\nabla \bm{u} - \bm{\varepsilon}_I) - \bm{f} (t)\cdot\bm{u} \right\rbrace
              - \int_{\partial \Omega_1} \bar{\bm{f}} (t)\cdot \bm{u}  
$$ 
The first term is the kinetic energy, $w$ is  the elastic strain energy, $\bm{f}$ is the volume force and $\bar{\bm{f}}$ is the surface force on the part $\partial \Omega_1$ of the boundary, the displacement field being equal to an imposed value $\bar{\bm{u}}$ on the remaining part $\partial \Omega_0$. 

Its symplectic gradient, according to Definition \ref{dsg},  is
$$ X H = ((D_{\bm{p}} H,D_{\bm{\pi}} H), (- D_{\bm{u}} H, - D_{\bm{\varepsilon}_I} H))
$$
where, introducing as usual the stress field 
$$ \bm{\sigma} = D w (\nabla \bm{u} - \bm{\varepsilon}_I)
$$
$D_{\bm{u}} H$ is the gradient in the variational sense  (from Definition \ref{dsg} and the integral form of the duality product)
$$ D_{\bm{u}} H = H_{,\bm{u}} - \nabla \cdot (H_{,\nabla\bm{u}})
                = - \bm{f} - \nabla \cdot \bm{\sigma}
$$
and
$$ D_{\bar{\bm{u}}} H = \bm{\sigma}\cdot\bm{n} - \bar{\bm{f}}
$$
Thus one has
$$ \dot{z}_I = \dot{z} - X H = \left(\left(\dot{\bm{u}} - \frac{\bm{p}}{\rho},\dot{\bm{\varepsilon}}_I \right),
          \left(\dot{\bm{p}} - \bm{f} - \nabla \cdot \bm{\sigma},\dot{\bm{\pi}} - \bm{\sigma}\right)\right)
$$

We shall use a dissipation potential which has an integral form: 
$$\Phi(z) \, = \, \int_{\Omega} \phi(\bm{p}, \bm{\pi}) \mbox{ d} x$$ 
and we shall assume that the symplectic Fenchel transform of $\Phi$ expresses as the integral of the symplectic Fenchel transform of the dissipation potential density $\phi$. 

The symplectic Fenchel transform of the function $\phi$ reads
$$\phi^{*\omega} (\dot{z}_I) 
        = \sup \left\{ \left\langle \dot{\bm{u}}_I, \dot{\bm{p}}'\right\rangle
                     + \left\langle \dot{\bm{\varepsilon}}_I, \dot{\bm{\pi}}'\right\rangle
                     - \left\langle \dot{\bm{u}}', \dot{\bm{p}}_I\right\rangle
                     - \left\langle \dot{\bm{\varepsilon}}'_I, \dot{\bm{p}}_I\right\rangle
                     - \phi(\dot{z}') \mbox{ : } \dot{z}' \in X \times Y \right\} 
$$
To recover the standard plasticity, we suppose that $\phi$ is depending explicitly only on $\dot{\bm{\pi}}$
$$ \phi(\dot{z}) = \varphi (\dot{\bm{\pi}})
$$
Denoting by $\chi_K $ the indicator function of a set $K$  (equal to $0$ on $K$ and to $+\infty$ otherwise), we obtain
$$\phi^{*\omega} (\dot{z}_I) = \chi_{\left\lbrace\bm{0}\right\rbrace} (\dot{\bm{u}}_I)
                             + \chi_{\left\lbrace\bm{0}\right\rbrace} (\dot{\bm{p}}_I)
                             + \chi_{\left\lbrace\bm{0}\right\rbrace} (\dot{\bm{\pi}}_I)
                             + \varphi^* (\dot{\bm{\varepsilon}}_I)
$$
where $\varphi^*$ is the usual Fenchel transform.  In other words,  the quantity $\displaystyle \phi^{*\omega} (\dot{z}_I)$ is finite if and only if all of the following  are true: 
\begin{enumerate}
\item[(a)] $\displaystyle \phi^{*\omega} (\dot{z}_I) = \varphi^* (\dot{\bm{\varepsilon}}_I)
$ , 
\item[(b) ] $\bm{p}$ equals  the linear momentum 
\begin{equation}
   \bm{p} = \rho \dot{\bm{u}}
\label{linear momentum defi} 
\end{equation}
\item[(c)] the balance of linear momentum is satisfied
\begin{equation}
   \nabla \cdot \bm{\sigma} + \bm{f} = \dot{\bm{p}} = \rho \ddot{\bm{u}}\quad \mbox{on} \quad\Omega,\qquad
                                       \bm{\sigma}\cdot\bm{n} = \bar{\bm{f}} \quad \mbox{on}\quad  \partial \Omega_1
\label{balance of linear momentum} 
\end{equation}
\item[(d)] and an equality which reveals the meaning of the variable $\pi$: 
\begin{equation}
   \dot{\bm{\pi}} = \bm{\sigma} \quad . 
\label{pi defi} 
\end{equation}
\end{enumerate}

The symplectic BEN principle applied to standard plasticity states that the evolution curve minimize: 
\begin{equation}
\Pi(z) = \int_{0}^{T} \left\{\varphi (\dot{\bm{\pi}}) + \varphi^* (\dot{\bm{\varepsilon}}_I)  -  \frac{\partial H}{\partial t}(t,z) \right\} \mbox{ dt} +  H(T, z(T)) 
\label{sbenp}
\end{equation}
among all curves $z: [0,T] \rightarrow X \times Y$ such that $\displaystyle z(0) = (x_{0}, y_{0})$, the kinematical conditions on $\partial \Omega_0$, (\ref{linear momentum defi}), (\ref{balance of linear momentum}) and (\ref{pi defi}) are satisfied.

\paragraph{The symplectic BEN principle and the original Brezis-Ekeland-Nayroles principle.}  
 Let us examine the important case where the kinetic energy and inertia forces can be neglected (quasi-static behaviour)
$$ \dot{\bm{p}} = \bm{0},\qquad
    H (t, z) = \int_{\Omega} \left\lbrace w(\nabla \bm{u} - \bm{\varepsilon}_I) - \bm{f} (t)\cdot\bm{u} \right\rbrace
              - \int_{\partial \Omega_1} \bar{\bm{f}} (t)\cdot \bm{u}  
$$ 
and the elasticity is linear
$$ \dot{\bm{\varepsilon}}_I = \nabla \dot{\bm{u}} - \bm{S} \dot{\bm{\sigma}}
$$
denoting $ \bm{S} = (D w )^{-1}$ the compliance operator. Eliminating $\bm{\pi}$ and $\bm{p}$ thanks to (\ref{linear momentum defi}) and (\ref{pi defi}), the symplectic BEN   principle (\ref{sbenp}) is transformed and claims that, the curve $\bm{u}: [0,T] \rightarrow U $ satisfying the kinematical conditions on $\partial \Omega_0$ being given, the evolution curve minimizes: 
\begin{equation}
\Pi(\bm{\sigma}) = \int_{0}^{T} \left\{\varphi (\bm{\sigma}) + \varphi^* (\nabla \dot{\bm{u}} - \bm{S} \dot{\bm{\sigma}})  - \frac{\partial H}{\partial t}(t,z) \right\} \mbox{ dt} + H(T, z(T)) 
\label{sbenp2}
\end{equation}
among all curves $\bm{\sigma}: [0,T] \rightarrow  E$ such that $\bm{\sigma} (0) = \bm{\sigma}_0$  and
\begin{equation}
   \nabla \cdot \bm{\sigma} + \bm{f} = \bm{0} \quad \mbox{on} \quad\Omega,\qquad
                                       \bm{\sigma}\cdot\bm{n} = \bar{\bm{f}} \quad \mbox{on}\quad  \partial \Omega_1
\label{equilibrium} 
\end{equation}
are satisfied. This expression can be transformed as follows. For sake of easiness, let us put:
$$ \langle \bm{l} (t), u \rangle = \int_{\Omega} \bm{f} (t)\cdot\bm{u} 
              + \int_{\partial \Omega_1} \bar{\bm{f}} (t)\cdot \bm{u}  
$$ 
Then, 
$$ \frac{\partial H}{\partial t}(t,z) = - \langle \dot{\bm{l}} (t), \bm{u} \rangle
$$
In the other hand
$$\frac{d}{dt} \left[ H(t,z(t)) \right] = \langle \bm{\sigma}, \nabla \dot{\bm{u}} - \dot{\bm{\varepsilon}}_I \rangle
                    - \langle \bm{l} (t), \dot{\bm{u}} \rangle - \langle \dot{\bm{l}} (t), u \rangle
$$
For the minimizer, the kinematical conditions on $\partial \Omega_0$  and the equilibrium equations (\ref{equilibrium}) are satisfied and using Green's formula:
\begin{equation}
\langle \bm{\sigma}, \nabla \dot{\bm{u}}  \rangle
                    = \langle \bm{l} (t), \dot{\bm{u}} \rangle
\label{Green formula} 
\end{equation}
that leads to
$$ \frac{d}{dt} \left[ H(t,z(t)) \right] - \frac{\partial H}{\partial t}(t,z) = - \langle \bm{\sigma},  \dot{\bm{\varepsilon}}_I \rangle
$$
Time-integrating and replacing in (\ref{sbenp2}) gives:
$$ \Pi(\bm{\sigma}) = \int_{0}^{T} \left\{\varphi (\bm{\sigma}) + \varphi^* (\nabla \dot{\bm{u}} - \bm{S} \dot{\bm{\sigma}})  - \langle \bm{\sigma}, \nabla \dot{\bm{u}} - \bm{S} \dot{\bm{\sigma}}  \rangle\right\} \mbox{ dt} + H (0, z (0))
$$

Integrating the latter term of the integral and forgetting the constant leads to the original Brezis-Ekeland-Nayroles principle \cite{Brezis Ekeland 1976} \cite{Nayroles 1976}. 
The evolution curve minimizes: 
\begin{equation}
\bar{\Pi}(\bm{\sigma}) = \int_{0}^{T} \left\{\varphi (\bm{\sigma}) + \varphi^* (\nabla \dot{\bm{u}} - \bm{S} \dot{\bm{\sigma}})  - \langle \bm{\sigma}, \nabla \dot{\bm{u}} \rangle\right\} \mbox{ dt} 
 + \dfrac{1}{2}\,\langle \bm{\sigma} (T), \bm{S} \bm{\sigma} (T) \rangle
\label{sbenp3} 
\end{equation}
among all curves $\bm{\sigma}: [0,T] \rightarrow  E$ such that $\bm{\sigma} (0) = \bm{\sigma}_0$  and the equilibrium equations (\ref{equilibrium}) are satisfied.

\section{Concluding remarks and future work}

It is worth to remark that in general the displacement evolution is also an unknown of the structure problem. This suggests considering another variant formulation relaxing the equilibrium equations. Using once again Green's formula (\ref{Green formula}) for the minimizer leads to claim that the evolution curve minimizes:
\begin{equation}
\bar{\Pi}(\bm{u},\bm{\sigma}) = \int_{0}^{T} \left\{\varphi (\bm{\sigma}) + \varphi^* (\nabla \dot{\bm{u}} - \bm{S} \dot{\bm{\sigma}})  - \langle \bm{l} (t), \dot{\bm{u}} \rangle \right\} \mbox{ dt} 
 + \dfrac{1}{2}\,\langle \bm{\sigma} (T), \bm{S} \bm{\sigma} (T) \rangle
\label{sbenp4} 
\end{equation}
among all curves $(\bm{u},\bm{\sigma}): [0,T] \rightarrow  U \times E$ such that $\bm{\sigma} (0) = \bm{\sigma}_0$  and the kinematical conditions on $\partial \Omega_0$ are satisfied. The variation with respect to the stress field allows to recover the dissipative constitutive law while the one with respect to the velocity field allows to recover the equilibrium equations. This space-time variational principle turns out to be a powerful alternative to the classical step-by-step approaches in the sense it works on the whole evolution of the system in the spirit of Ladev\`{e}ze's LATIN method \cite{Ladeveze 1985}, \cite{Boisse 1989}, \cite{Boisse 1990}, \cite{Boisse 1991}, \cite{Ladeveze 1991}. It averts typical pitfalls of step-by-step approaches which accumulate errors steps after steps and may fail in case of non convergence at a given step.

In the future, we wish to develop the applications of this new theoretical formalism according to the three objectives mentioned in the introduction: extending to the dissipative systems the geometrical methods of classical dynamics by suitable change of coordinates (in particular using canonical transformations, Lie group theory and momentum maps), exploring the dissipative rheological models in dynamical situations and using dynamical Brezis-Ekeland-Fenchel principle to solve evolution problems by suitable numerical algorithms. Also, we aim to generalize this approach to implicit standard materials by introducing a symplectic bipotential.


\begin{thebibliography}{10}


\bibitem{Ammar 2006} A. Ammar, B. Mokdad, F. Chinesta, R. Keunings, A new family of solvers for some classes of multidimensional partial differential equations encountered in kinetic theory modeling of complex fluids, {\it J. Non-Newtonian Fluid Mech.},{\bf 139},153-176, 2006.

\bibitem{Ammar 2007} A. Ammar, B. Mokdad, F. Chinesta, R. Keunings, A new family of solvers for some classes of multidimensional partial differential equations encountered in kinetic theory modeling of complex fluids,  Part II: Transient simulation using space-time separated representations, {\it J. Non-Newtonian Fluid Mech.},{\bf 144},98-121, 2007.

\bibitem{ambtor} L. Ambrosio, V. Tortorelli, On the Approximation of Free Discontinuity Problems, |{\it Bollettino UMI} {\bf 7},  6-B, 105-123, 1992.

\bibitem{aubin2} J.-P. Aubin, Boundary-Value Problems for Systems of Hamilton-Jacobi-Bellman Inclusions with Constraints, {\it SIAM J. Control}, {\bf 41}, 425-456, 2002.

\bibitem{aubin} J.-P. Aubin, A. Cellina, J. Nohel,  Monotone trajectories of multivalued dynamical systems, {\it Annali di Matematica Pura ed Appl.}, {\bf 115}, 99-117, 1977.

\bibitem{bloch} A.M. Bloch, P.S. Krishnaprasad, J.E. Marsden, T.S. Ratiu, Dissipation induced instabilities, {\it Ann. de l'Institut Henri Poincar\'e. Analyse non lin\'eaire}, {\bf 11}, 1, 37-90, 1994.

\bibitem{bodo} G. Bodovill\'e, On damage and implicit standard materials,  {\it C. R. Acad.  Sci. Paris S\'erie IIB}, {\bf 327}, 8, 715-720, 1999.

\bibitem{bodo sax EJM 01} G. Bobovill\'e, G. de Saxc\'e, Plasticity with non linear kinematic hardening : modelling and shakedown analysis by the bipotential approach, {\it Eur. J. Mech. A/Solids}, {\bf 20}, 99-112, 2001.

\bibitem{Bognet 2012} B. Bognet, F. Bordeu, F. Chinesta, A. Leygue, A. Poitou, Advanced simulation of models defined in plate geometries: 3D solutions with 2D computational complexity, {\it Comput. Meth. in Appl. Mech. and Eng.}, {\bf 201-204}, 1-12, 2012.

\bibitem{Boisse 1989} P. Boisse, P. Ladev\`{e}ze, P. Roug\'ee, A large time increment method for elastoplastic problems, {\it Eur. J. Mech. A/Solids}, {\bf 8}, 257-275, 1989.

\bibitem{Boisse 1990} P. Boisse, P. Bussy, P. Ladev\`{e}ze, A new approach in non-linear mechanics: the large time increment method, {\it Int. J. Num. Meth. Eng.}, {\bf 29}, 647-663, 1990.

\bibitem{Boisse 1991} P. Boisse, P. Ladev\`{e}ze, M. Poss, P. Roug\'ee, A new large time increment algorithm for anisotropic plasticity, {\it Int. J. Plasticity}, {\bf 7}, 65-77, 1991.

\bibitem{Brezis Ekeland 1976} H. Brezis and I. Ekeland, Un principe variationnel associ\'e \`{a} certaines \'{e}quations paraboliques. I. Le cas ind\'ependant du temps, II. Le cas d\'ependant du temps. {\it C. R. Acad. Sci. Paris S\'erie A-B}, 282, 971-974, and 1197-1198, 1976.

\bibitem{bham} M. Buliga, Hamiltonian inclusions with convex dissipation with a view towards applications, {\it Mathematics and its Applications} {\bf 1}, 2 (2009), 228-251, \href{http://arxiv.org/abs/0810.1419}{arXiv:0810.1419}.

\bibitem{dangsax} K. Dang Van, G. de Saxc\'e, G. Maier, C. Polizzotto, A. Ponter, A. Siemaszko, D. Weichert, Inelastic Behaviour of Structures under Variable Repeated Loads. D. Weichert G. Maier, Eds., CISM International Centre for Mechanical Sciences, Courses and Lectures, {\bf 432}, Springer, 2002.

\bibitem{Fitzpatrick 1988} Fitzpatrick, S, Representing monotone operators by convex functions. In: Workshop/Miniconference on Functional Analysis and Optimization, Canberra, 1988, pp. 59–65. Proc. Centre Math. Anal. Austral. Nat. Univ., 20, Austral. Nat. Univ., Canberra, 1988.

\bibitem{Giner Chinesta 2013} E. Giner, B. Bognet, J. J. R\'{o}denas, A. Leygue, F. Javier  Fuenmajor, F. Chinesta, The proper Generalized Decomposition (PGD) as a numerical procedure to solve 3D cracked plates in linear fracture mechanics, {\it Int. J. Solids and Structures}, {\bf 50}, 1710-1720, 2013.

\bibitem{hjiaj bodo CRAS 00} M. Hjiaj, G. Bobovill\'e, G. de Saxc\'e, Mat\'eriaux viscoplastiques et loi de normalit\'e implicites, {\it C. R. Acad. Sci. Paris S\'erie IIb}, {\bf 328}, 519-524, 2000.

\bibitem{Ladeveze 1985} P. Ladev\`{e}ze, Sur une famille d'algorithmes en M\'ecanique des Structures, {\it C. R. Acad. Sci. Paris S\'erie II}, {\bf 300}, 41-44, 1985.

\bibitem{Ladeveze 1991} P. Ladev\`{e}ze, New advances in large time increment method. In P. Ladev\`{e}ze, and O.C. Zienkiewicz, eds., {\it New advances in computational structural mechanics}, 3-21, Elsevier, 1991.

\bibitem{mielke} A. Mielke, Evolution in rate-independent systems (Ch. 6). In C. Dafermos, E. Feireisl, eds., {\it Handbook of Differential Equations, Evolutionary Equations, vol. 2}, 461-559, Elsevier, 2005.

\bibitem{MR06b} A. Mielke, T. Roub\'{\i}\v{c}ek, Rate-independent damage processes in nonlinear elasticity, {\it Mathematical Models and Methods in Applied Sciences (M3AS)}, {\bf 16}, 2, 177-209, 2006.

\bibitem{mielketh99} A. Mielke, F. Theil. A mathematical model for rate-independent phase transformations with hysteresis. In H.-D. Alber, R. Balean, and R. Farwig, editors, Proceedings of the Workshop on Models of Continuum Mechanics in Analysis and Engineering, 117-129. Shaker-Verlag, 1999.

\bibitem{mielkethl} A Mielke, F. Theil, V. Levitas,  A Variational Formulation of Rate-Independent Phase Transformations  Using an Extremum Principle, {\it Archive for Rational Mechanics and Analysis}, {\bf 162}, 2, 137-177, 2002.

\bibitem{Nayroles 1976} B. Nayroles, Deux th\'eor\`{e}mes de minimum pour certains syst\`{e}mes dissipatifs, {\it C. R. Acad. Sci. Paris S\'erie A-B}, 282, A1035-A1038, 1976.

\bibitem{rocka} R.T. Rockafellar, Generalized Hamiltonian equations for convex problems of Lagrange, {\it Pacific J. of Math.}, {\bf 33}, no. 2, 411-427, 1970.

\bibitem{saxfeng} G. de Saxc\'e, Z.Q. Feng, New inequation and functional for contact with friction : the implicit standard material approach, {\it Int. J. Mech. of Struct. and Machines}, {\bf 19}, 3,  301-325, 1991.

\bibitem{sax CRAS 92} G. de Saxc\'e, Une g\'en\'eralisation de l'in\'egalit\'e de Fenchel et ses applications aux lois constitutives, {\it C. R. Acad. Sci. Paris S\'erie II}, {\bf 314}, 125-129, 1992. 

\bibitem{sax boussh IJMS 98} G. de Saxc\'e, L. Bousshine, Limit Analysis Theorems for the Implicit Standard Materials: Application to the Unilateral Contact with Dry Friction and the Non Associated Flow Rules in Soils and Rocks,  {\it Int. J. Mech. Sci.}, {\bf 40}, 4, 387-398, 1998.

\bibitem{vall leri CONST 05} C. Vall\'ee, C. Lerintiu, D. Fortun\'e, M. Ban, G. de Saxc\'e, A bipotential expressing simultaneous ordered spectral decomposition between stress and strain rate tensor. International conference New Trends in Continuum Mechanics, Constanta (Romania), September 8-12, 2003, published under the title "Hill's bipotential", in New Trends in Continuum Mechanics, Ed. Theta, 339-351, 2005.

\bibitem{visintin} A. Visintin, Structural stability of rate-independent nonpotential flows, {\it Discrete and Continuous Dynamical Systems Series S}, {\bf 6}, 257-275, 2013.

\bibitem{Zouain 2010} Zouain, N., Pontes Filho, I., Vaunat, J., Potentials for the modified Cam-Clay model. {\it European Journal of Mechanics A/Solids}, {\bf 29}, 327-336, 2010.


\end{thebibliography}
\end{document}